\documentclass[preprint,prc,aps,showpacs,preprintnumbers,amsmath,aps,amssymb]{revtex4-1}
\usepackage{graphicx}
\usepackage{dcolumn}
\usepackage{bm}
\usepackage{color}
\usepackage{afterpage}

\begin{document}

\title{Nuclear evaporation process with simultaneous multiparticle emission}
\author{Leonardo P. G. De Assis}
\email{lpgassis@cbpf.br}
\author{S\'{e}rgio B. Duarte}
\email{sbd@cbpf.br}
\affiliation{ Centro Brasileiro de Pesquisas F\'isicas\\
Rua Dr. Xavier Sigaud 150, 22290-180 Rio de Janeiro--RJ, Brasil}%
\author{Bianca M. Santos}
\email{biankce@cbpf.br}
\affiliation{Instituto de F\'isica, Universidade Federal Fluminense\\
Av. Gal. Milton Tavares de Souza, 24210-346 Niter\'oi--RJ, Brasil}%

\date{\today}
\begin{abstract}
The nuclear evaporation process is reformulated by taking into account simultaneous multiparticle emission from a hot compound nucleus appearing as an intermediate state in many nuclear reaction mechanisms. The simultaneous emission of many particles is particularly relevant for high excitation energy of the compound nucleus.These channels are effectively open in competition with the single particle emissions and fission in this energy regime. Indeed, the inclusion of these channels along the decay evaporating chain shows that the yield of charged particles and occurrence of fission are affected by these multiparticle emission processes of the compounded nucleus, when compared to the single sequential emission results. The effect also shows a qualitative change in the neutron multiplicity of different heavy  compound nucleus considered. This should be an important aspect for the study of spallation reaction in Acceleration Driven System (ADS) reactors. The majority of neutrons generated in these reactions come from the evaporation stage of the reaction, the source of neutron for the system. A Monte Carlo simulation is employed to determine the effect of these channels on the particle yield and fission process. The relevance of the simultaneous particle emission with the increasing of excitation energy of the compound nucleus is explicitly shown.
\end{abstract}
 \pacs{ 24.10.Pa, 24.60.Dr, 25.40.Sc, 25.85.-w}
 \maketitle
 \newpage

\section{INTRODUCTION}

Nuclear evaporation is the mechanism process usually employed for a hot
intermediate compound nucleus formed in a nuclear reaction. When a target nucleus is  bombarded by a massive particle or photon, a rapid cascade phase
occurs with ejection of few nucleons, leading to the formation of a hot residual
nucleus. At this point, the evaporation of particles begins through an
evaporative decay chain. The fission reaction of the nucleus in the chain is
considered in competition with the emission of neutrons, protons and alpha
particles.

Presently, a great interest has been dedicated to this type of reaction,
which is used to describe the neutron production for the newest proposed
generation of nuclear reactors, the Accelerator Driven System (ADS)
\cite{rubbia}-\cite{anefalos2008}. The system is composed of a particle
accelerator coupled to a reactor. The accelerated beam of charged particles
collides with nuclei in the reactor medium, generating neutrons by spallation
reactions. This neutron generation process works as an external source to
induce fission in the reactor cycle. The main interest for this device arises
due to it operates in a sub-critical regime, burning radioactive
nuclear waste originated in conventional machines. However, as a relatively
new proposal, and involving high energy beams, the ADS requires a deeper
study, particularly concerning the spallation reaction.

For a typical beam energy of ADS system (around 1 GeV for proton
beam), the formed compound nucleus can reach an excitation energy of a few
hundred MeV, which is high enough to permit not only single emission, but also
simultaneous particle evaporation\textbf{.} Such simultaneous emission
channels constitute an open issue to be explored.
The evaporation reaction chain includes many nuclei far from the beta
stability line with no experimental data available. The nuclear properties of
these nuclei are the object of intense theoretical and experimental physical
research in order to understand the structure of these exotic nuclei.
Consequently, in some calculations we deal with reasonable working hypothesis or extrapolated results extracted from the conventional model.

The changes in the conventional evaporation process in order to include multiparticle simultaneous emission is described in the next section and the details of the 
channel widths calculation is presented in section III. The implementation of computational Monte Carlo simulation in the new version of the evaporation phase is shown in section IV. The results of our calculation and comparison with the conventional one is depicted in section V, followed by our concluding remarks in the last section.

\section{EVAPORATION PROCESS: SINGLE EMISSION X SIMULTANEOUS MULTIPLE EMISSION}

In a conventional treatment of nuclear evaporation, particles mainly neutron, proton and alpha are emitted sequentially along an evaporative decay chain. The decay sequence is ended by the fission process or by the exhaustion of the residual nucleus excitation energy. In usual Weisskopf prescription to the nuclear evaporation \cite{weiss37}, the probabilities of sequential single particle emission or fission are given by \cite{weiss37,bianca}:%

\begin{equation}
w_{\lambda}=\frac{\Gamma_{\lambda}}{\underset{\lambda}{\sum}\Gamma_{_{\lambda
}}}\
\end{equation}

\noindent with $\Gamma_{\lambda}$ being the width of the $\lambda$-channel of single emission (proton, neutron, alpha particle) or the fission channel. After the particle emission the excitation energy of the residual nucleus decreases once the particle kinetic energy is carried away from the nuclear system,    

\begin{equation}
E_{f}^{\ast}=E^{\ast}-\overline{E}_{ev}.
\end{equation},

\noindent where $\overline{E}_{ev_{k}}= \overline\epsilon_{i} + S_{i}$, being $\epsilon_{i}$ and $S_{i}$ the average kinetic energy of the emitted nucleon and its separation energy, respectively.

Fission processes are considered in competition with single sequential particle emissions, and whenever the fission process occurs, the decay chain is ended. When it does not occur the decay chain proceeds until the excitation energy is exhausted with a final spallation residual nuclear being formed. 

The multiple simultaneous emission of particles are introduced to take into account the emission of particles dragged by outgoing energetic evaporated particles. This can be induced by a short range of correlations between the group of neighbour particles due to residual interactions between them, particularly important at region near the nuclear surface. The correlation is not maintained outside the nuclear medium and the group soon disintegrates outside the nucleus. In this work, we disregard binding energy for the emitted set of particles wich is an important aspect to determine the energy carried out from the hot nucleus in each emission process. However, there are many conceptual difficulties that must be overcome to incorporate these decay channels consistently once scarce literature is founded on simultaneous multiparticles evaporation mechanism \cite{tomasini57}-\cite{Gazis}. One of the pioneering work on this issue is the statistical Tomasini's study \cite{tomasini57}, where N-particle phase space is used as a microstate representation of the emitted particles. After little more than half century of publication, this pioneering study remains a current work on this subject. Essentially, the calculations are based on fundamental ideas still valid nowadays. Some years later, another evaporation approach with a modified calculation for one-particle density level was introduced by the work of Beard et al \cite{beard63}. In spite of getting more realistic nuclear particle density level distribution, the Woods-Saxon nuclear potential form was used instead of a potential well, and the work still use sequential particle emission in the evaporating decay chain. We are analysing the effect of the multiparticle emission mechanism by using a prescription similar to that the one employed by Tomasini approach, but making it in accordance with Weisskopf treatment for the conventional case of single emission. 

\begin{figure}[ht]
\centering
\resizebox*{0.52\textwidth}{!}{\rotatebox{0}{
\includegraphics{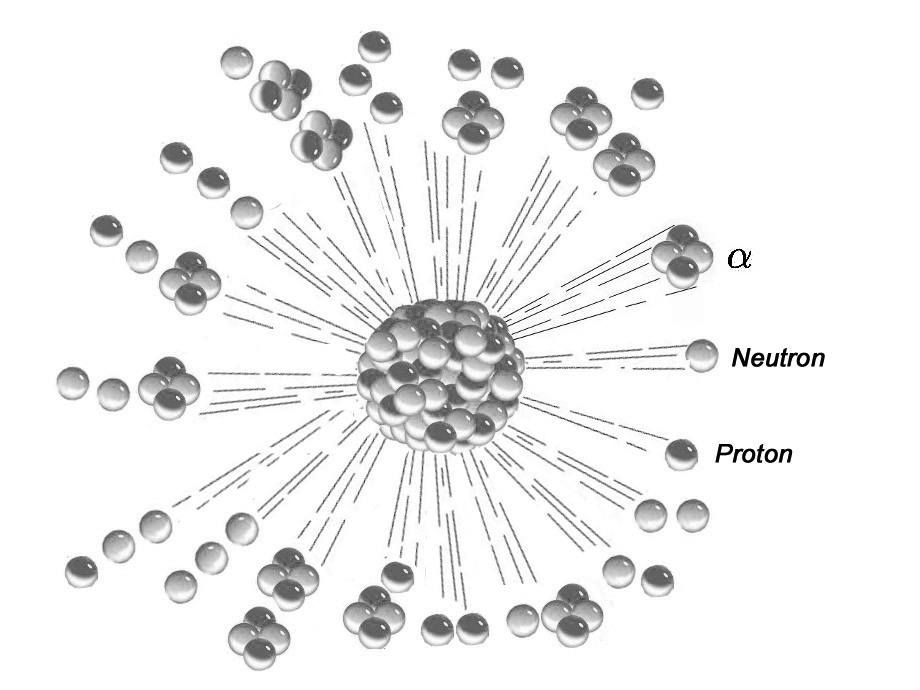}}}
\vspace{0.2cm}\caption{Different channels considered in the evaporate decay chain
taking into account simultaneous multiparticle emission processes. 
\label{fig:fig1}}
\end{figure}

Many particle simultaneous emission channels implemented in the new evaporative decay chain are illustrated in Fig.\ref{fig:fig1}. In each radial direction, it is indicated a different channel. The probability density of multiple simultaneous emission is determined in connection with the Fermi Gold Rule \cite{fermi50} for the transition rate of the process as,

\begin{equation}
W=g_{1,2,...,N}\left\vert H_{0}\right\vert ^{2}\varrho_{N}%
V^{N}\varpi(E_{f}^{\ast}),\label{mu1}%
\end{equation}

\noindent where $E_{f}^{*}=E^{*}-\epsilon _{ev} -S_{N}$, with  $S_{N}=\sum\limits_{i=1}^{N}S_{i}$ and  $\epsilon_{ev}=\sum\limits_{i=1}^{N}\epsilon_{i}$. The factor $g_{123..N}$ is the spin factor associated to the emitted $N$-particles in the decay channel, and $|H_{0}|^{2}$ is the square of transition matrix element. The number of microstates in the N-particle phase space is $\rho_{N}V^{N}$, with

\begin{equation}
\varrho_{N}=\frac{\left(  m_{1}m_{2}...m_{N}\right)  ^{\frac{3}{2}}}%
{2^{\frac{3}{2}N}\pi^{\frac{3}{3}N}\hbar^{N}}\frac{\varepsilon^{\frac{3}%
{2}N-1}}{\left(  \frac{3}{2}N-1\right)  !},\label{mu2}%
\end{equation}

\bigskip

\noindent where it was used a non-relativistic dispersion relation. The final state of the remaining nucleus should be determined by the specification of the function $\varpi\left(E_{f}^{*},Z_{f}^{*},A_{f}^{*}\right)$ in Eq. (\ref{mu1}).

\bigskip

 For the case of single particle emission, Tomasini's approach should be reduced to Weisskopf treatment. By taking this recognition procedure, the function $\varpi$ can be identified and used in other cases.  From eq.(\ref{mu1}), expressions of  density of probability are prepared focusing the most relevant decays channels (one, two and three simultaneous particle emissions) to be used in our calculation. Thus, for these channels we can write,   

\begin{equation}
W_{1}\left(  \epsilon\right)  d\epsilon=Kg_{1}\frac{4A}{3\sqrt{2\pi}
}\frac{m_{1}^{\frac{3}{2}}}{\mu^{3}}\epsilon^{\frac{1}{2}}\varpi\left(
E^{*}-\epsilon-S_{1}\right)  d\epsilon,
\label{eq:mu5}
\end{equation}

\bigskip

\begin{equation}
W_{1,2}\left(  \epsilon\right)  d\epsilon=Kg_{12}\frac{A^{2}}{9\pi}
\frac{\left(  m_{1}m_{2}\right)  ^{\frac{3}{2}}}{\mu^{6}}\epsilon^{2}
\varpi\left(  E^{*}-\epsilon-S_{1}-S_{2}\right)  d\epsilon,
\label{eq:mu6}
\end{equation}

\bigskip

\begin{equation}
W_{1,2,3}\left(  \varepsilon\right)  d\varepsilon=Kg_{123}\frac{32\sqrt{2}A^{3}
}{2835\pi^{2}}\frac{\left(  m_{1}m_{2}m_{3}\right)  ^{\frac{3}{2}}}{\mu^{9}
}\varepsilon^{\frac{7}{2}}\varpi\left(  E^{*}-\varepsilon-S_{1}-S_{2}
-S_{3}\right)  d\varepsilon.
\label{eq:mu7}
\end{equation}

\bigskip

 In the expressions above $\mu$ represents the pion mass,  used to determine the distance scale of the nucleon-nucleon interaction by the uncertainty relation.

 From the expression of the single particle emission density of probability in Eq.(\ref{eq:mu5}), we can compare it with Weisskopf's result (see, for example Ref.\cite{Cole} ), and we obtain the explicit form of the function associated to the nuclear states in Tomasini's approach. Namely,

\begin{equation}
\varpi\left(  E^{\ast}-\varepsilon-S_{N}\right)  =\rho_{f}\left(  E^{\ast
}_{f}\right)  /\rho\left(  E^{\ast}\right).
 \label{mu31}
\end{equation}

\bigskip

 The constant $K$ in Eqs. (\ref{eq:mu5}-\ref{eq:mu7}) will be eliminated later in determining the branching ratio of decay channels. Disregarding this constant $K$, the probability of each channel can be obtained by  integrating  the above density of probability for all values of the total kinetic energy of the N-particles, 

\begin{equation}
W=\int\limits_{0}^{E-B}W_{i}\left(  \varepsilon\right)  d\varepsilon
\label{mu8}
\end{equation}

\noindent with $\varepsilon=\sum\limits_{i=1}^{N}\varepsilon_{i}$. After the integration, the following expressions are obtainned,

\begin{equation}
\mathit{W}_{1}=\frac{2}{3}{\frac{Ag_{{i}}\sqrt{2}{m_{{i}}}^{3/2}
\sqrt{\mathit{\varepsilon}}\rho_{{f}}\left(  E^{{*}
}_{f}\right)  }{\pi\,{\mu}^{3}{\mathit{\rho}}_{{i}
}\left(  E^{*}\right)  }} 
\label{mu22}
\end{equation}

\bigskip

\begin{equation}
\mathit{W}_{\mathit{12}}=\,\frac{1}{9}{\frac{{A}^{2}g_{1}g_{2}\left(  {m}%
_{{1}}{m_{2}}\right)  ^{3/2}{\mathit{\varepsilon}}^{2}\rho_{{f}}\left(
E^{*}_{f}\right) }{\pi\,{\mu}^{6}%
\mathit{\rho}_{{i}}\left(  E^{*}\right)  }}\label{mu23}%
\end{equation}

\bigskip%

\begin{equation}
\mathit{W}_{123}={\frac{32}{2835}}\,{\frac{{A}^{3}g_{1}g_{{2}}g_{3}\sqrt
{2}\left(  {m_{1}m_{2}m_{3}}\right)  ^{3/2}{\mathit{\varepsilon}}%
^{7/2}\rho_{{f}}\left(  E^{*}_{f}\right)
}{{\pi}^{2}{\mu}^{9}\mathit{\rho}_{{i}}\left(  E^{*}\right)  }}.\label{mu24}%
\end{equation}

\bigskip%

 We can rewrite the density of nuclear states in terms of the entropy,\ $\eta(E)=\ln\left(\rho\left(E \right)\right) $, and by using the relationship between entropy and temperature $\frac{d\eta}{dE}=\frac{1}{T(E)}$ and making use of the same approximation currently employed in the Weisskopf's approach, the following result is obtained for the correspondent width of channels      

\begin{equation}
\Gamma_{1}=\frac{1}{3}\,\frac{\hslash\sqrt{2}Ag_{{1}}{m_{{1}}}^{3/2}%
{\mathrm{e}^{-2\ \,\sqrt{aE^{{*}}}+2\,\sqrt{a\left(  \ \ E^{{*}%
}-V\ \right)  \ \ }}}\left(  {\frac{E^{{*}}-V}{a}}\right)  ^{3/4}%
}{{\mu}^{3}\sqrt{\pi}},\label{mu34}%
\end{equation}

\bigskip%

\begin{equation}
\Gamma_{12}=\frac{2}{9}\,\frac{\hslash{A}^{2}g_{1}g_{2}\left(  {m_{1}m_{2}%
}\right)  ^{3/2}{\mathrm{e}^{-2\,\sqrt{aE^{{*}}}+2\,\sqrt{a\left(
\ \ E^{{*}}-V\ \right)  \ \ }}}\left(  {\frac{E^{{*}}%
-V}{a}}\right)  ^{3/2}}{{\pi\mu}^{6}},\label{mu35}%
\end{equation}

\bigskip%

\begin{equation}
\Gamma_{123}={\frac{2}{27}}\,\frac{\hslash{A}^{3}g_{1}g_{2}g_{{3}}\left(
{m_{1}m_{2}m_{3}}\right)  ^{3/2}\sqrt{2}{\mathrm{e}^{-2\,\sqrt{\ aE^{{*}}%
}+2\,\sqrt{a\left(  \ \ E^{{*}%
}-V\ \right)  \ \ }}}\left(  {\frac{E^{{*}}-V}{a}}\right)  ^{9/4}%
}{{\pi}^{3/2}{\mu}^{9}},\label{mu36}%
\end{equation}

\noindent where $a$ is the parameter of nuclear density level. In the above expressions of channel width, the  relationship between  the excitation energy of the nuclear state and the effective nuclear temperature is employed, $(E^{*}=T^{2}/a)$.  Moreover, it was considered the Fermi gas model to determine the entropy of the initial compound nucleus and of the residual one. The determination of the effective coulomb barrier $V$ and values of the density of level parameters are detailed in the next section.

\section{MONTE CARLO SIMULATION OF EVAPORATION CHAIN WITH MULTIPARTICLES EMISSION CHANNELS}

A Monte Carlo calculation is employed to simulate the decay chain by sampling
the channel among the whole set of considered decay processes (single,
multiple emissions and fission channels). When fission occurs, the evaporation chain is ended. After each step of particle emission, the remaining excitation energy
of the nucleus is determined and a new branch of the chain is created. If no
energy is available for the subsequent decay, neither fission nor evaporation
can take place. In this process, spallation nuclear product is formed and the chain is
terminated. The number of different emitted particles along the chain is
accumulated to generate the particle yield. The number of times that the
$i-th$ channel is accessed generates the channel access number along the
evaporation process.

After each decay step, to determine the remaining excitation energy of the residual nucleus  we have to consider the energy carried away by the
emitted set of particles, through the separation energies of them. When charged particles are emitted (protons and alphas) the effective Coulomb barrier is taken into account. Thus, the emission treated in each step of the decay chain, the potential barrier is determined as,

\begin{equation}
V=CK_{\lambda}\frac{Z_{\lambda}(Z-Z_{\lambda})e^{2}}{r_{0}(A-A_{\lambda})^{1/3}+r_{0}
A_{\lambda}^{1/3}}, 
\label{eq:barr}
\end{equation}

\noindent where the subscript $\lambda$ specifies the channel with the nuclear radius parameter taken as $r_{0}=1.2$ fm. 

The effective barrier penetrability factor, $K_{\lambda}$, for the channel including $n$-protons and $m$-alphas, is given approximately by,

\begin{equation}
K_{\lambda}=K_{p}^{n}K_{\alpha}^{m},
\end{equation}

\noindent where $K_{p}=0.78$ and $K_{\alpha}=0.83$ are the proton and alpha barrier
penetrability, respectively.

By having the tunnelling process well defined, the internal degree of freedom of the outgoing N-particle system is considered frozen. For this particular purpose, the whole set is taken as a particle, as considered in the Weisskopf's treatment(see \cite{Cole}). 

The equation (\ref{eq:barr}), $C=1-E^{\ast}/B(Z,A)$ represents the thermal
correction to the Coulomb barrier \cite{tavares1992} of the
decaying nucleus with $B(Z,A)$ being its binding energy of the nucleus before the emission.

The mean energy removed by evaporated particles from nucleus in this decay step is calculated as \cite{tavares2004},

\begin{equation}
\overline{E}_{ev}=\left(  S_{N}+V\right)  +2\sqrt{\frac{E^{\ast}-\left(
S_{N}+V\right)  }{a}}.
\end{equation}

\noindent where $S_{N}$ is the sum of emitted particles separation energy, and 
$V$ is the potential barrier for the monopole coulomb interaction between whole group of particles and residual nucleus. The density level parameter $a$ is obtained by the Fermi gas approximation for a nuclear system, parametrized as in Ref.\cite{tavares2004} and considered approximately the same for different emitted particles. The nuclear density level parameter used is given by,     

\begin{equation}
a=\tilde{a}\left\{  1+\left[  1-\exp\left(  -0.05E^{\ast}\right)  \right]
\frac{\Delta M}{E^{\ast}}\right\} , \label{eq:m1}
\end{equation}

\noindent where $\Delta M$ (in $MeV$) is the shell correction extracted from
the mass formula \cite{Myers77} and

\begin{equation}
\widetilde{a}=0.114A+0.098A^{\frac{2}{3}}MeV^{-1}
\end{equation}

\noindent is a systematic parametrization for the asymptotic
value of level density parameter.

\bigskip

 The fission width is given by \cite{Nix}%

\begin{equation}
\Gamma_{f}=\frac{\left[  2a_{f}^{1/2}\sqrt{E^{\ast}-B_{f}}-1\right]  }{4\pi
a_{f}\exp\left[  2\sqrt{a_{c}E^{\ast}}\right]  }\exp\left[  2a_{f}^{1/2}%
\sqrt{E^{\ast}-B_{f}}\right]
\end{equation}

\bigskip

\noindent where $a_{f}$ and $a_{c}$ are saddle point and compound\ nucleus density parameters respectively \cite{Nix}. The height of the barrier fission process is given by $B_{f}=C\cdot B_{f0}$ \cite{tavares1992} with $C$ the thermal correction appearing in Eq. (\ref{eq:barr}), here applied to the  cold fission barrier,   $B_{f0}=\Delta m-\Delta M$.  The excess of masses $\Delta m$ are obtained from Ref \cite{Myers77} and the $\Delta M$ are the shell corrections to nuclear masses as used in Eq (\ref{eq:m1}).

\bigskip The calculation of the density level parameter for fission process
($a_{f}$) received a different treatment because of the shell effects of
fission barrier. The value of $a_{f}$ was obtained by $r\cdot a$, in wich
the value of $r$ is extracted by the general expression \cite{Tavares2004b}:%

\begin{equation}
r=1+\frac{p\left(  A\right)  }{E^{\ast q\left(  A\right)  }}%
\end{equation}

For excitation energies greater than about $40$ MeV, the values {}{}of $p$ and
$q$ are given by the following expressions:  

\bigskip%

\begin{align*}
p\left(  A\right)   &  =\exp\left[  0.150\left(  222-A\right)  \right]  \text{
\ \ \ }150\leq A\leq210,\\
p\left(  A\right)   &  =\exp\left[  0.257\left(  217-A\right)  \right]  \text{
\ \ \ }210\leq A\leq232,\\
q\left(  A\right)   &  =0.0352\left(  235-A\right)  \text{ \ \ \ \ \ \ \ \ \ }%
150\leq A\leq232.
\end{align*}

Now, with all the widths of considered $\lambda $-channel calculated, the corresponding decay branch ratio is determined, $\left.  \Gamma_{\lambda}\right/{\displaystyle\sum}\Gamma_{\lambda}$. Here $\lambda$ is covering single, double and triple emission of particles. This branch ratio is used to sort the decay channel to be followed in each step of the decaying chain.  

Along the calculation, the type of emitted particles and the occurrence of fission process, as well as the residual nucleus of the process are recorded.
Through the recorded data, particles yields and compound nucleus fissility can be calculated.

\section{RESULTS}

\begin{center}
\begin{figure}[ht]
\resizebox*{0.9\textwidth}{!}{\rotatebox{0}
{\includegraphics[width=145mm]{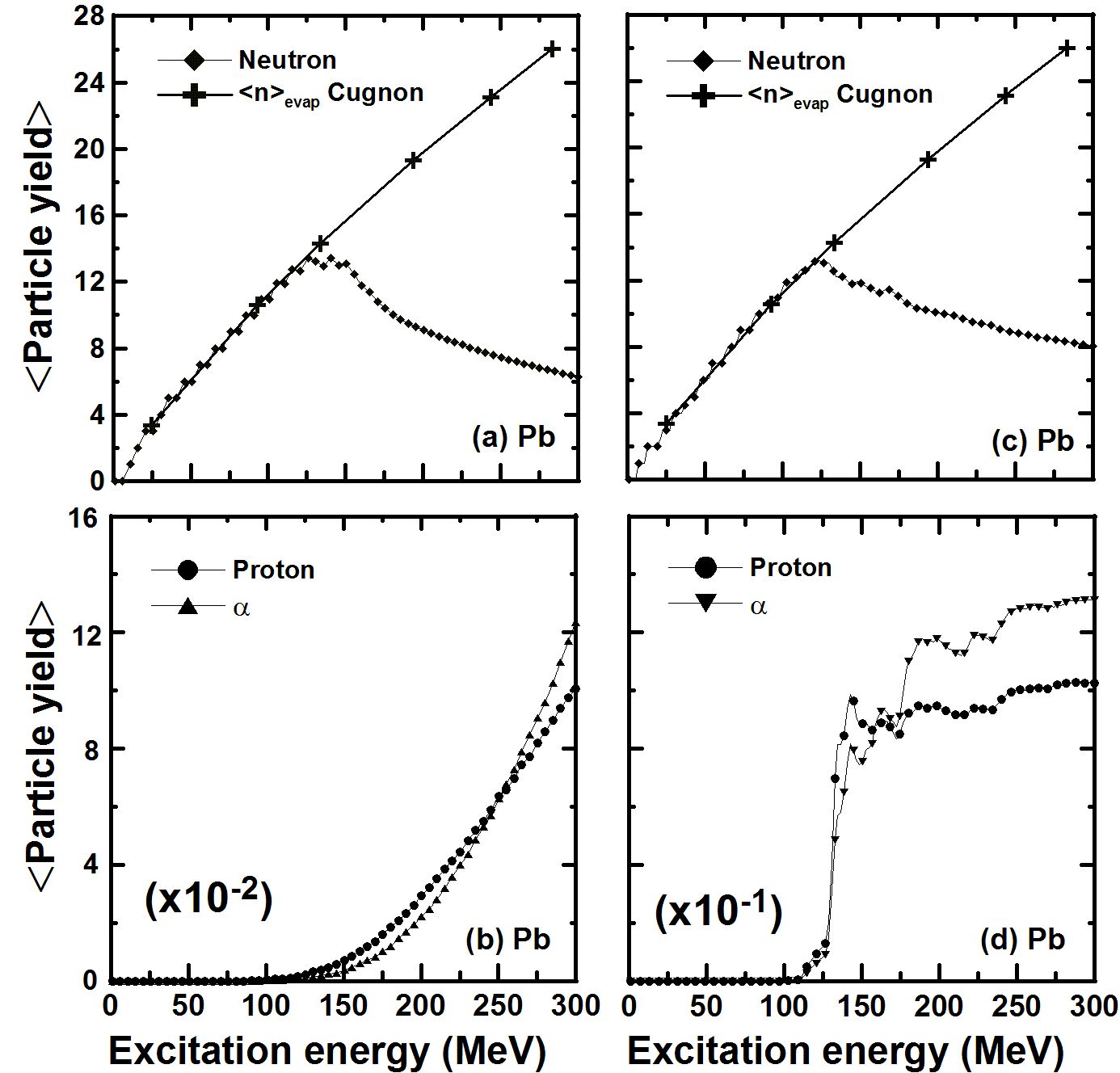}}}
\vspace{0.2cm}\caption{ Yield of evaporated particles as a function of excitation energy for $^{208}$Pb nucleus. The results of calculations using only single sequential emission (in parts-(a) and (d)) is compared with that ones from  calculation including simultaneous multiparticle emission (in parts (b) and (d) ). The solid curve in parts (a) and (c) are the neutron yield resulting from conventional calculation in Ref. \cite{Cugnon}. Note that the parametrization is for the yield of evaporated particle disregarding the competition with fission channel (included in our result). 
\label{fig:fig2}}
\end{figure}
\end{center}

The calculation of the new evaporative chain was carried out by running 600 thousand Monte Carlo repetitions to obtain statistical final results. $^{208}$Pb and $^{200}$Hg are taken as representative nuclei to show the difference between evaporative process with and without the inclusion of multiparticle emission channels. As a first result, in Fig.\ref{fig:fig2} we compare the total particle yield considering only sequential single emission for $^{208}$Pb. Parts (a) and (c) of the shown results of the calculation using only sequential single particle emissions and parts (b) and (d) corresponds to the same calculation including the new muliparticle channels. We can see that the neutron yield for the sequential single emission in part-(a) presents a well located maximum around $150$ MeV, which is slightly lower than  in part-(b, followed by a little steeply decreasing tail. At this energy the included channels begin a small contribution to the neutron yield, once single emission is still the dominant process when including multiparticle emission. Results of our calculation coincides with conventional evaporation process results till this energy. The comparison with the results of the Liege group in Ref.\cite{Cugnon} is show by the solid curve in parts (a) and (b). Comparing the yield of the emitted charged particles (protons and alpha) we note one change of one order of magnitude favouring the results of calculation including multiparticle emission. In addition, a qualitative change in the curve behavior is observed comparing part (b) and (d).

\begin{center}
\begin{figure*}[ht]
\resizebox*{1.0\textwidth}{!}{\rotatebox{0}{\includegraphics[width=145mm]{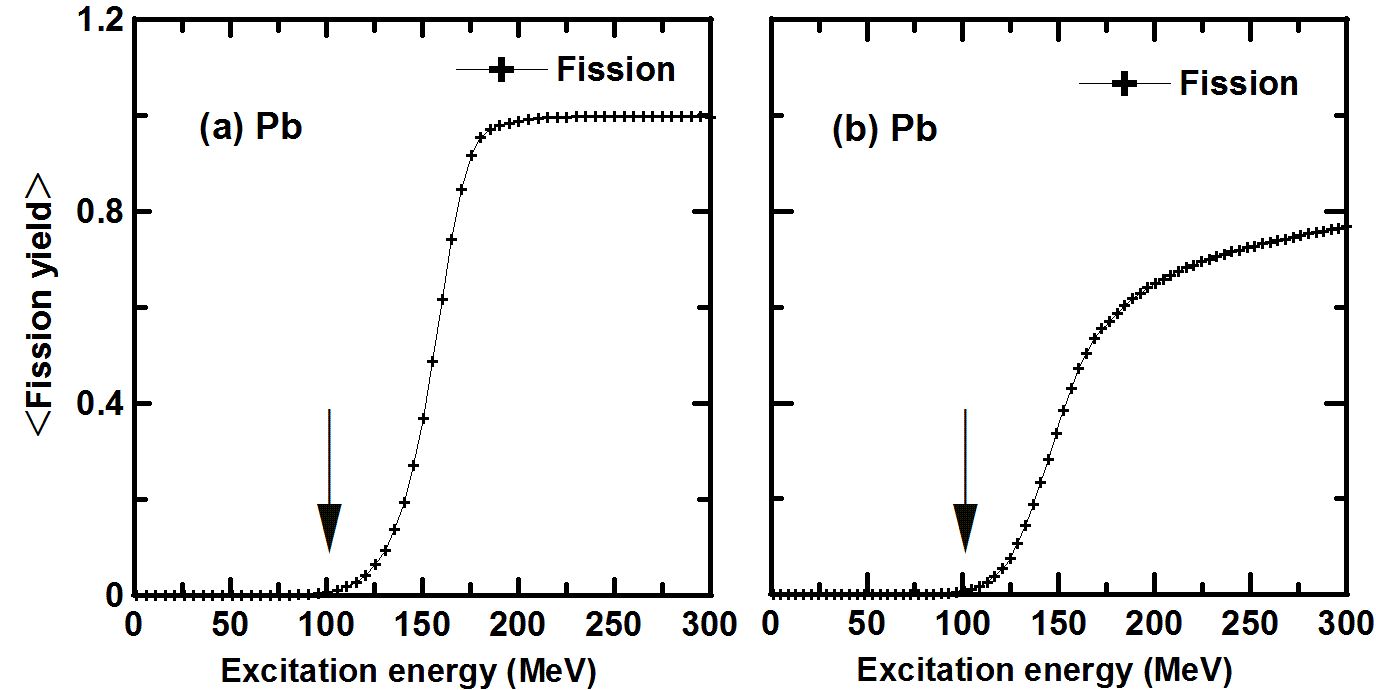}}}
\vspace{0.2cm}\caption{Yield of fission process for the case of single
sequential emission calculation in part-a and for simultaneous
multiparticle emission case, in part-b, for $^{208}$Pb. The arrow indicates the threshold energy for fission process along the decay chain.
\label{fig:fig3}}%
\end{figure*}
\end{center}

In Fig.\ref{fig:fig3}, we compare the average number of fission events for the calculations for single sequential particle emission ( part-(a)) with the results including multiparticle emission (part-(b). The result are similar, however the competition between new channels turns down the occurrence of fission process significantly at high excitation energy regime, as we can see in part-(b) of Fig.\ref{fig:fig3}. The arrow indicates the similar threshold of excitation energy for both type of calculation.

\begin{figure}[htb]
\centering
\resizebox*{0.9\textwidth}{!}{\rotatebox{0}{\includegraphics[width=145mm]{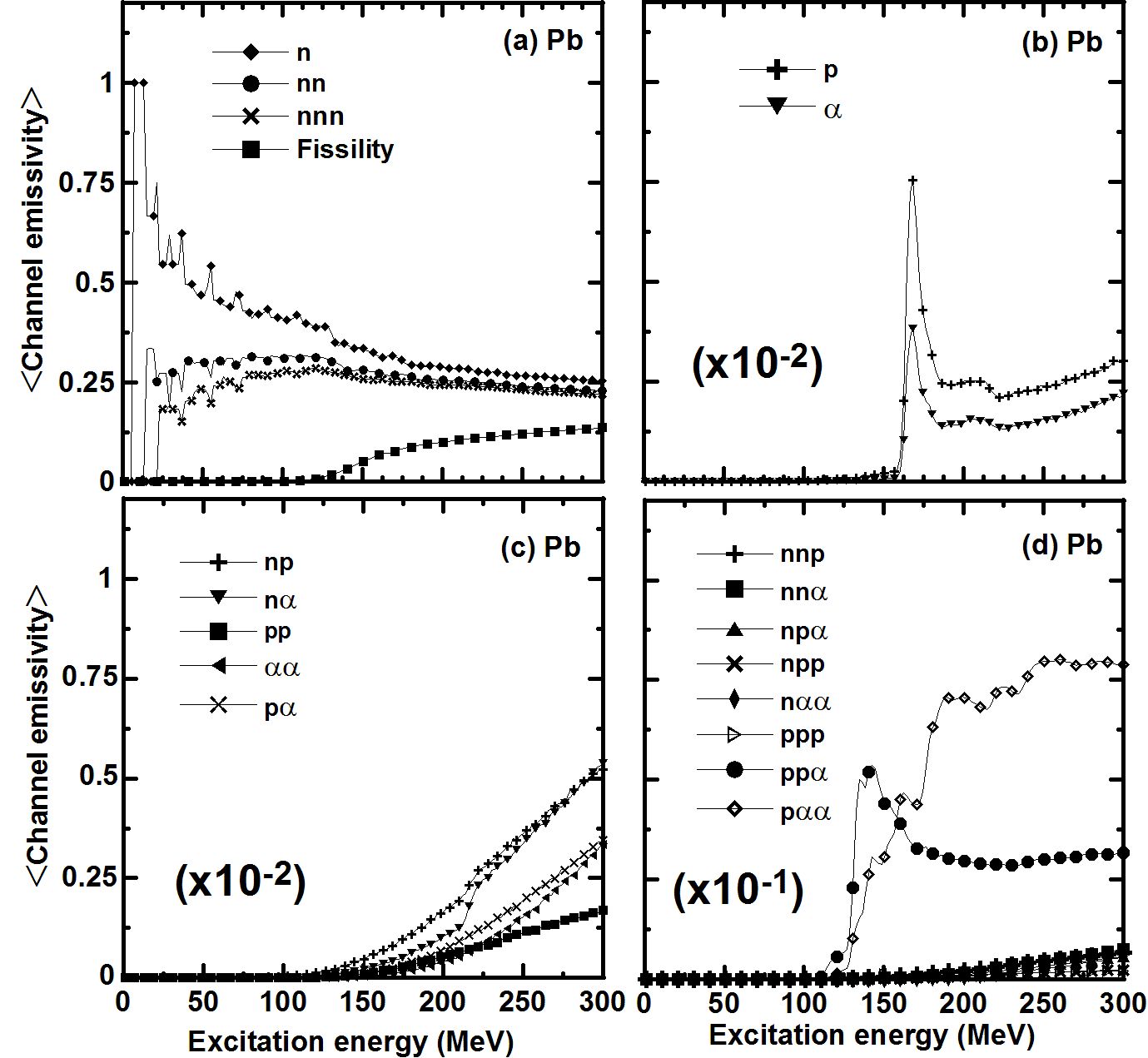}}}
\vspace{0.2cm}
\caption{Channel branching ratios for the decay of $^{208}$Pb excited nucleus. They are grouped by similar magnitude of the average number of accesses. In part-(a) the dominant channels are grouped, the neutron emission (single, double and ternary) and fission processes. The single emission of charged particles (proton and alpha)are grouped in part-(b). The binary and ternary simultaneous emissions are in part-(c) and part-(d), respectively. In brackets the  multiplicative factor for vertical scales (in parts (b), (c) and (d) ).
\label{fig:fig4}}%
\end{figure}

The competition level of the different channels can be observed in Fig \ref{fig:fig4}, showing the branching ratio of the whole set of included channels as a function of compound nucleus excitation energy.  Single emission of charged particles are grouped in part-(b) and ternary emission including charged particles in part-(d). In fact, the branching of some ternary emission in part-(d) $p p \alpha$ and $p \alpha \alpha$  are one order of magnitude greater than single and double particle emission including charged particles (parts (b) and (c) ). These are the channel contribution for the significant yield of protons and alphas in part-(d) of Fig \ref{fig:fig2}. 

\begin{figure}[ht]
\centering
\resizebox*{0.9\textwidth}{!}{\rotatebox{0}{\includegraphics[width=145mm]{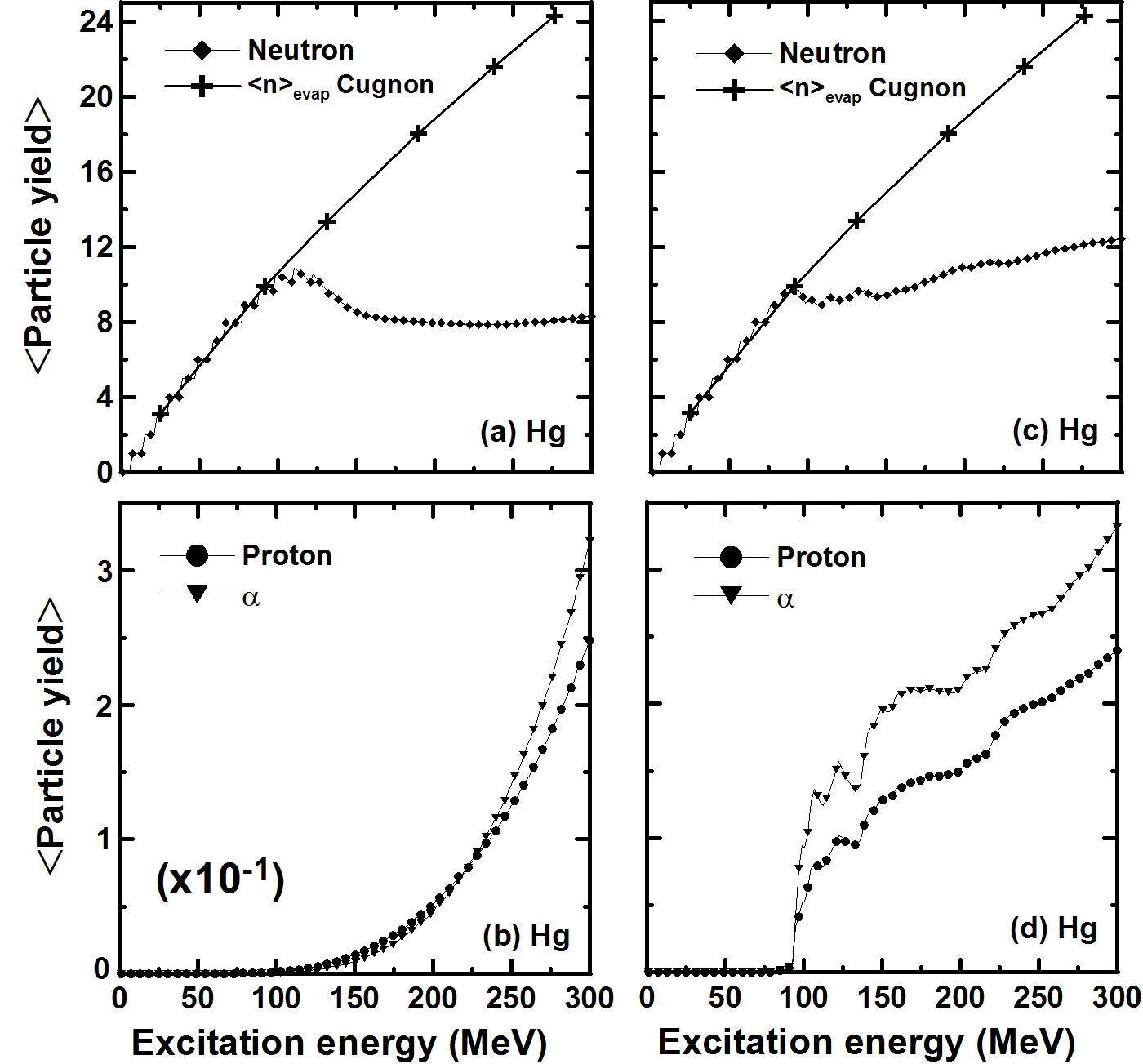}}}
\caption{ Yield of evaporated particles as a function of excitation energy for $^{200}$Hg nucleus. The results of calculations using only single sequential emission (in parts-(a) and (d)) is compared with that ones from  calculation including simultaneous multiparticle emission (in parts (b) and (d) ). The solid curve in parts (a) and (c) are the neutron yield resulting from results of conventional calculation in Ref. \cite{Cugnon}. Note that their parametrization is for the yield of evaporated particle disregarding the competition with fission channel (included in our result). 
\label{fig:fig5}}
\end{figure}

\begin{figure}
\centering

\resizebox*{0.9\textwidth}{!}{\rotatebox{0}{\includegraphics[width=145mm]{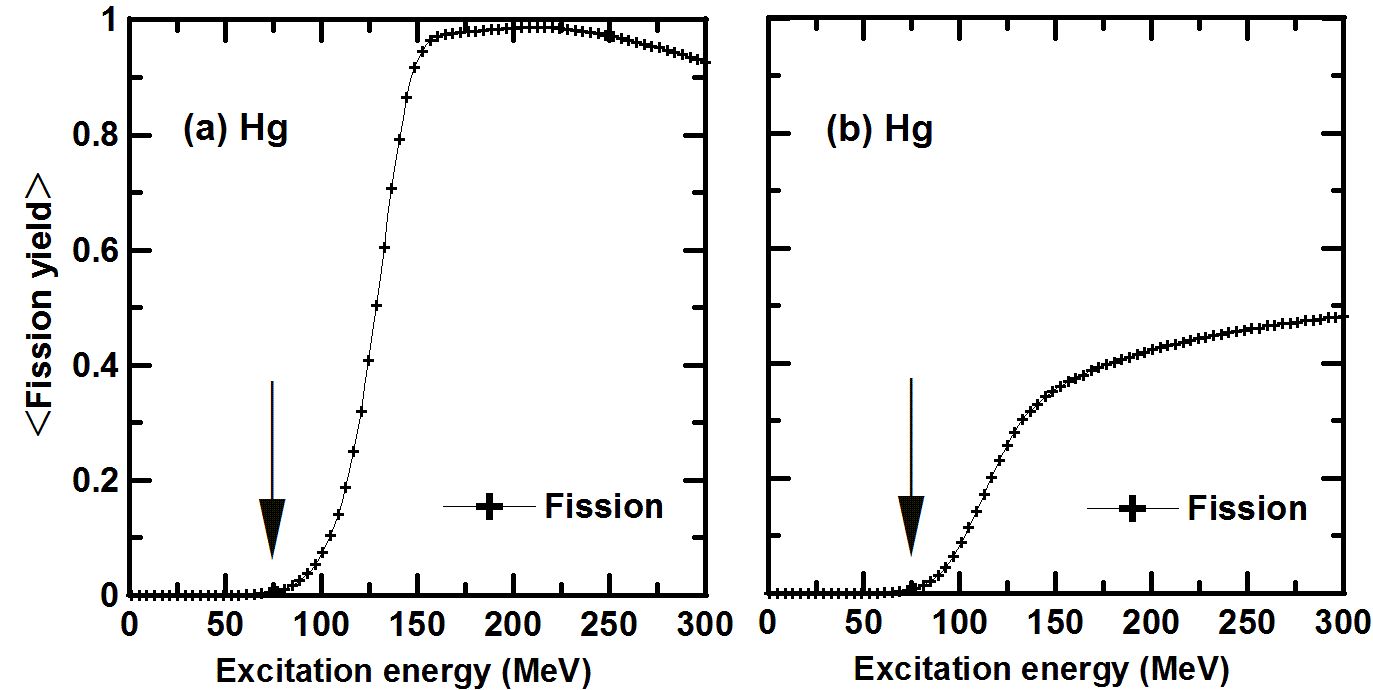}}}
\caption{Yield of fission along the decay chain of $^{200}$Hg for single sequential emission calculation (part-(a)) compared to simultaneous multiparticle calculation (part-(b). The arrow indicates the similar energy threshold for fission process occurring along of the decay chain. 
\label{fig:fig6}}
\end{figure}

\begin{figure}[ht]
\vspace{0.1cm} 
\centering
\resizebox*{0.9\textwidth}{!}{\rotatebox{0}{\includegraphics[width=145mm]{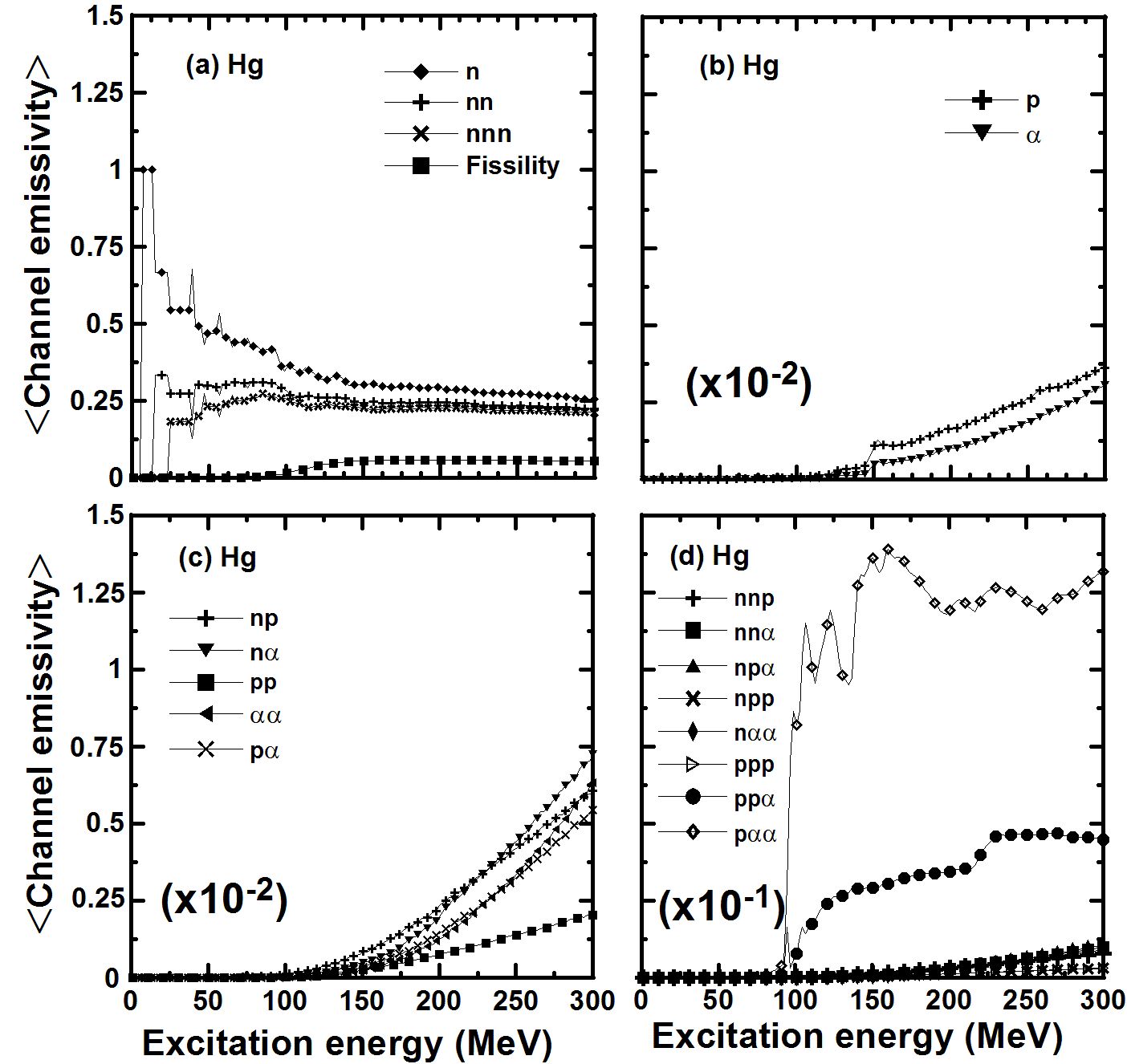}}}
\vspace{0.2cm}\caption{Channel branching ratios for the decay of $^{208}$Pb excited nucleus. They are grouped by similar magnitude of the average number of accesses. In part-(a) the dominant channels are grouped, the neutron emission (single, double and ternary) and fission processes. The single emission of charged particles (proton and alpha)are grouped in part-(b). The binary and ternary simultaneous emissions are in part-(c) and part-(d), respectively. In brackets the  multiplicative factor for vertical scales (in parts (b), (c) and (d) ).
\label{fig:fig7}}
\end{figure}

For the case of Hg, the results are shown in figures \ref{fig:fig5}-\ref{fig:fig7}. The comparison for the calculation using only single emission with multiparticle emission channels are shown in Fig.\ref{fig:fig5}. As observed, a small increasing in the neutron yield along the tail of curves in parts (a) and (c), more pronounced for the calculation including multiparcles emission channels in part-(c). In previous result for Pb, the curves in Fig.\ref{fig:fig2} exhibit a decreasing tail for higher excitation energy. The charged particle yields (proton and alphas) for the calculation including multiparticle emission channels are one order of magnitude higher than for the results obtained with only single sequential emission (this was also observed in the case of Pb). 

The Fig.\ref{fig:fig6} shows the average number of occurrences of fission events along the decay chain of the excited Hg compound nucleus. We can see a similar behaviour to that one for Pb shown in Fig.\ref{fig:fig3}, with a more pronounced reduction of the fission events for the case of multiple emissions and a slight decrease behavior for high excitation energy values.

In Fig.\ref{fig:fig7}, the branching ratio of the channels for $^{200}$Hg nucleus is shown. As in Fig.\ref{fig:fig4}, they are grouped by similar magnitudes. The dominant channels are those with only neutron emission (single, double and ternary) and fission processes, shown in part-(a). It is important to remark that in the case of Pb, the ternary simultaneous emissions $p p \alpha$ and $p \alpha \alpha$ in part-(d) are more significant than the single charged particle emission and double emission including charged particle shown in parts (b) and (c), respectively.

\bigskip

As in the case of Pb, the channels with only neutrons emission are the dominant ones, shown in part-(a). The remarkable point appears in part-(d), in which the ternary emission channels $p p \alpha$ and $p \alpha \alpha$ are more significant than the single emission of charged particles and double emission involving charged particle (This was also observed for the case of the Pb decay chain).

The main purpose of present work is to show the effect of multiparticles simultaneous emission channels on evaporating particles yields when we compare the traditional emission model with the case including multiple emission. Surely, it would be worthwhile to compare our results with some experimental data, however, the data for evaporating phase of reaction mechanism are not clean enough for this comparison. They are dependent on the experimental cut off energy separating particles from the compound nucleus evaporation as well as the low energy particles ejected in target cascade process. The use o thick targets in these experiments \cite{Leray, Letourneau} imposes a representative difficult. As a results, the experimental data offers only upper and lower bounds to our calculation of neutron yields.  The table \ref{experimental} confronts our results with experimental data (for thin and thick targets and energy of the proton in the reaction p + A of $800$, $1200$ , and $1600$ MeV). 

\pagebreak

\begin{table}[htb]
    \centering
\caption{Average multiplicities of neutrons emitted in the evaporation phase from reaction p+Pb experimental \cite{Leray, Letourneau}, compared with our results. Our result for the neutron multiplicity using only of single sequential evaporation emission is indicated by $<n_{s}>$, and for the case of simultaneous multiparticle emission by $<n_{m}>$. The experimental data \cite{Letourneau} shows results for total neutrons more likely emitted in evaporation phase.}
\vspace{0.4cm}
\label{experimental}
\begin{tabular}{lcccc}
\hline
             & \hspace*{2.0cm}  Calculated  &        &  \hspace*{2.0cm}        Data  &   \\\hline

  $E_{beam}$  \hspace*{0.8cm}           & $<n_{s}>   $   &\hspace*{0.8cm} $<n_{m}>$ \hspace*{0.6cm}      & Exp. up bound \hspace*{0.6cm}  & Exp. low bound  \\
\hline
$800 MeV$     &  $13.35$ & $13.08$      & $12.0\pm 6.57$  &   $6.5\pm 0.7$             \\
$1200 MeV$   &  $9.47$ & $10.23$ &    $12.12\pm 7.13$       & $8.3\pm 0.8$   \\
$1600 MeV$                &   $7.8$     & $9.16$          &    --          & $10.1\pm 1.0$           \\

\hline
\hline
 \\
\end{tabular}
\end{table}

\noindent Table I shows that our results are compatible with experimental limits data. 

\pagebreak

\section{CONCLUSION}

The evaporation of a hot compound nucleus is reconsidered taking into account 
inclusion of simultaneous multiple particle emission channels. The relevance
of these processes is shown for the case a heavy compound nucleus formed with
excitation energy in the range of few MeV to three hundred MeV. The average
particle yield is determined and compared to the case of the conventional
calculation. The relative significance of the channels is shown by calculating the channel branching ratio as a function of the excitation energy. Calculation for Pb and Hg are performed showing that the multiparticle emission  channels  are relevant for excitation energies higher than $150$ MeV. The result for neutron generation using multiple particle emissions calculation does not change  significantly the result of a conventional calculation. However, the  yield of charged particles and the fission occurrence along the compound nucleus decay chain are affected by the inclusion of the simultaneous multiple emission channels. This result may have implications on the fission and spallation fragments distributions in proton-nucleus reaction and  also on the yield of charged particle generated. This study can be investigated by  coupling the new evaporation code with the intranuclear cascade calculation  to simulate the whole process of spallation, which is a work in progress.

\section*{Acknowledgments}

 S. B. Duarte is grateful to CNPq Brazilian Research Agency for the invaluable financial support.
L.P.G. De Assis. is grateful to CNPq for his post-doctoral fellowship and Bianca M. Santos  expresses her gratitude to CAPES Brazilian Research Agency for her Ph.D. fellowship.

\end{document}